# Time-delay signature suppression in a chaotic semiconductor laser by fiber random grating induced distributed feedback

Yanping Xu,[1] Mingjiang Zhang,[1,2] Liang Zhang,[1] Ping Lu,[3] Stephen Mihailov,[3] and Xiaoyi Bao[1,*]

[1]*Department of Physics, University of Ottawa, 25 Templeton Street, Ottawa, Ontario K1N 6N5, Canada*

[2]*Institute of Optoelectronic Engineering, Department of Physics and Optoelectronics, Taiyuan University of Technology, Taiyuan 030024, China*

[3]*National Research Council Canada, 100 Sussex Drive, Ottawa, Ontario K1A 0R6, Canada*

## Abstract

We demonstrate that a semiconductor laser perturbed by the distributed feedback from a fiber random grating can emit light chaotically without the time delay signature. A theoretical model is developed based on the Lang-Kobayashi model in order to numerically explore the chaotic dynamics of the laser diode subjected to the random distributed feedback. It is predicted that the random distributed feedback is superior to the single reflection feedback in suppressing the time-delay signature. In experiments, a massive number of feedbacks with randomly varied time delays induced by a fiber random grating introduce large numbers of external cavity modes into the semiconductor laser, leading to the high dimension of chaotic dynamics and thus the concealment of the time delay signature. The obtained time delay signature with the maximum suppression is 0.0088, which is the smallest to date.

[*] Electronic mail: xiaoyi.bao@uottawa.ca.

Chaos theory reveals that a nonlinear high dimension system may bifurcate to more complex dynamics including chaos[1]. This erratic behavior has been investigated intensively in more practical lasers and has remained a fertile research field for the past decades[2,3]. Chaotic signals generated from semiconductor lasers are of great interests for a number of promising applications, including high-speed random bit generation, secure communication, data encryption, and chaotic ranging[4-15]. Particularly, laser diode based chaotic output has acted as a physical entropy source for true random number generation, benefiting from outcome unpredictability with no dependence on any previous outcome. The most effective and simplest method to perturb the single mode semiconductor laser into chaos is by returning a small fraction of the laser emission into the laser diode cavity using an external reflector[16-18]. Coherence collapse dynamics is expected when the frequency of the external cavity is much smaller than the laser's intrinsic relaxation oscillation frequency, leading to high-dimensional chaotic behavior[19]. However, such time-delay systems often suffer from feedback time delay signature (TDS), which could be easily identified by the autocorrelation function (ACF). TDS is potentially detrimental in chaos-based cryptosystems and truly random number generation[20]. Hence its suppression is important and highly desired for these applications. To this end, various feedback configurations, including simple single mirror feedback[21], dual-path feedback from double mirrors[22], and fiber Bragg grating feedback[23,24], have been investigated and studied. More complex configurations including mutually coupled feedback using two lasers[25,26], feedback with cascaded coupled laser injection[27], and on-chip integrated optical feedback[28] have also been proposed to suppress the TDS. However, these techniques increase the hardware complexity of the experimental setup and are not fully applicable in a practical sense.

In this paper, a simple approach for TDS suppression is developed by employing the feedback from a fiber random grating fabricated by the femtosecond (fs) laser with a large number ($>10^4$) of reflectors with random time delays. Randomly placed reflectors along the fiber provide effective random distributed feedback to a laser diode, which is capable of obscuring the TDS of the chaotic output. A theoretical model is established to illustrate the chaotic dynamics of the laser diode subjected to the random distributed feedback and the simulated results predict the larger suppression ratio than that of the single mirror feedback configuration. The experimental results obtained from the random grating feedback show the TDS could be suppressed to a value of 0.0088, which is the smallest value reported to date.

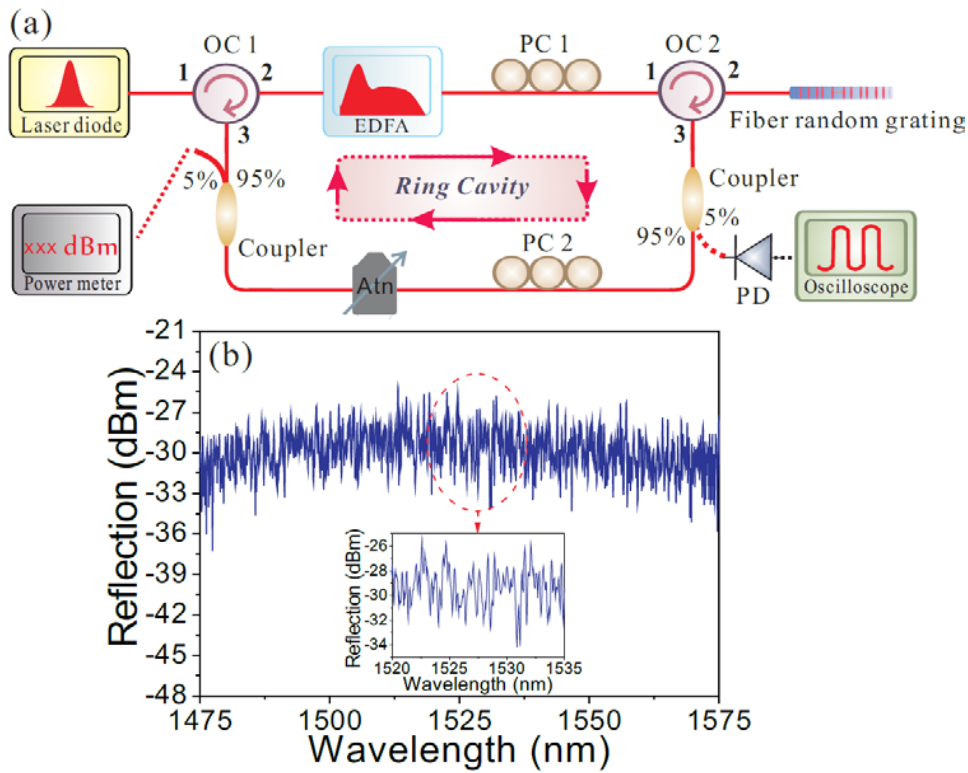

Figure 1. (a) Configuration of the laser diode subjected to fiber random grating feedback. OC: Optical Circulator; EDFA: Erbium Doped Fiber Amplifier; PC: Polarization Controller; Atn: Optical Attenuator; PD: Photo-detector; (b) Reflection spectrum of fiber random grating. Inset: enlarged reflection spectrum.

The configuration of the laser diode with random grating feedback is shown in Fig.1 (a). Light from a single mode laser diode was first amplified with an Erbium-doped fiber amplifier (EDFA) by passing it through an optical circulator (OC1). A polarization controller (PC1) was used to adjust the state of polarization (SOP) of the amplified light before it is reflected by the random grating. The fiber random grating was manufactured using a fs pulse duration regeneratively amplified Ti:sapphire laser operating at 800nm and having an 80 fs pulse duration. Index modification spots with random spacing from 0 to 3.5μm were written along the fiber axis using a precision air bearing stage to translate the fiber and 50× microscope objective to focus the beam[29]. Around 50000 spots were made along the 10cm long standard single mode fiber (SMF). Fig.1 (b) shows an example of the reflection spectrum of the random grating. A strong interference pattern generated by the backscattered light from different scattering centers is observed (see inset Fig.1(b)) and could be regarded as the result of the superposition of numerous Fabry-Perot interferometers and Mach-Zehnder interferometers formed by the index modification spots. The random distributed feedback from the random grating is then passed through a 95/5 coupler. The 5% of the tapped light was detected by a photo-detector (Newport) with a bandwidth of 25GHz and its time domain series was recorded by an oscilloscope. The 95% of the tapped light was sent as the feedback to the laser diode. PC2 was to ensure that the polarizations reflected back from the random grating were equal with that entering into. An optical attenuator was used to adjust the strength of the feedback to the laser diode and a power meter was used to measure the feedback strength.

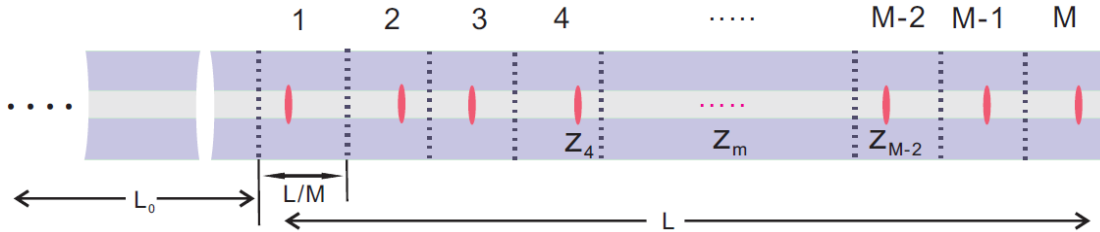

Figure 2. Simplified model of the randomly distributed scattering centers in fiber random grating.

In order to investigate the influence of random distributed optical feedback on the dynamics of the chaotic semiconductor laser, the fiber random grating could be theoretically modeled as a collection of large number of enhanced Rayleigh scattering centers randomly distributed along the SMF with L=10cm. Each scattering center introduces optical feedback with a specific delay to the laser diode. As shown in Fig. 2, a simplified model is developed for the random grating sample. The random grating can be divided into M sections with equal length L/M. Each section has one scattering center, which is randomly located within the section with uniform distribution and assumed to have equal reflection coefficients. The position of each scattering center is denoted as $Z_m$. The optical path length of the fiber ring structure is 53.5m and the leading fiber length of the random grating sample is $L_0$=4.01m, corresponding to a total time delay of 307.6ns.

If the light from the laser diode is denoted as $E(t)=A(t)e^{i\varphi(t)}$, under the random distributed optical feedback and neglecting multiple reflections among the scattering centers, the laser field can be numerically simulated by modifying the Lang-Kobayashi model[20, 30-33] as follows:

$$\frac{dA}{dt} = \frac{1}{2}\left[\frac{g(N-N_0)}{1+\varepsilon|E|^2} - \frac{1}{\tau_p}\right]A + \frac{1}{\tau_{in}}\sum_{m=1}^{M}k_m A(t-\tau_m)\cos[\omega\tau_m + \varphi(t) - \varphi(t-\tau_m)] + F_A(t)$$

$$\frac{d\varphi}{dt} = \frac{1}{2}\alpha\left[\frac{g(N-N_0)}{1+\varepsilon A^2} - \frac{1}{\tau_p}\right] + \frac{1}{\tau_{in}A}\sum_{m=1}^{M}k_m A(t-\tau_m)\sin[\omega\tau_m + \varphi(t) - \varphi(t-\tau_m)] + F_\varphi(t) \quad (1)$$

$$\frac{dN}{dt} = pJ_{th} - \frac{N}{\tau_N} - \frac{g(N-N_0)}{1+\varepsilon A^2}A^2$$

with

$$F_A(t) = 2\beta N/A + \sqrt{2\beta N}\zeta_A(t)$$
$$F_\varphi(t) = (1/A)\sqrt{2\beta N}\zeta_\varphi(t) \quad (2)$$

where $k_m$ and $\tau_m$ are the feedback reflectivity and feedback delay for each scattering center in the fiber random grating, respectively. $k_m$ is equal for all of the M scattering centers and $\tau_m$ follows uniform distribution within each divided section. Considering the computing efficiency and the limited computing power of the computers in the lab, simulations with M=100 and 200 were performed, respectively. Although the M value used in simulation is smaller than the number of reflectors in the fiber random grating, it is still able to predict the TDS suppression as shown later in this letter. The parameters of the laser used in simulations are listed as follows: $N_0$=4.55×10$^{23}$ m$^{-3}$ is the transparency carrier density, $g$=3×10$^{-12}$ m$^3$s$^{-1}$ is the differential gain, $\varepsilon$=5×10$^{-23}$ m$^3$ is the gain saturation parameter, $\tau_p$=1.17ps is the photon lifetime, $\tau_N$=2.5ns is the carrier lifetime, $\alpha$=5.0 is the linewidth enhancement factor, $\tau_{in}$=7.38ps is the round-trip time in the laser diode cavity, $J_{th}$=6.568×10$^{32}$ m$^{-3}$s$^{-1}$ is the threshold current density, $p$=2.0 is the pumping factor, $\omega=2\pi c/\lambda$ is the angular frequency of the laser field, where $\lambda$=1550nm and $c$=3×10$^8$ m/s, $F_A(t)$ and $F\varphi(t)$ are polar decomposition in amplitude and phase of the spontaneous emission noise, including the fluctuations arising from the spontaneous process and denoted in the form of the Langevin noise source, $\beta$ is the spontaneous emission rate, $\zeta_A(t)$ and $\zeta_\varphi(t)$ represent uncorrelated white Gaussian noises with zero mean and correlations following $<\zeta_A(t)>=<\zeta_\varphi(t)>=0$ and $<\zeta_A(t)\zeta_A(t')>=<\zeta_\varphi(t)\zeta_\varphi(t')>=\delta(t-t')$ and $<\zeta_A(t)\zeta_\varphi(t')>=0$, the spontaneous emission noise is about eight orders of magnitude smaller than the laser field in amplitude.

Numerical simulations were performed using Eq. (1) and (2) with random distributed feedback and single reflection feedback, respectively. In simulations, the current density $J$ was set as $2J_{th}$, at which the intrinsic relaxation oscillation frequency of the laser diode was around 7GHz. The ACF of the chaotic laser output is computed to obtain the TDS using the following equation :

$$C(\tau) = \frac{\langle (I(t)-\langle I(t)\rangle)(I(t+\tau)-\langle I(t)\rangle)\rangle}{(\langle (I(t)-\langle I(t)\rangle)^2\rangle \langle (I(t+\tau)-\langle I(t)\rangle)^2\rangle)^{1/2}} \tag{3}$$

where $I(t)$ is the chaotic time domain series, $\tau$ is the time delay.

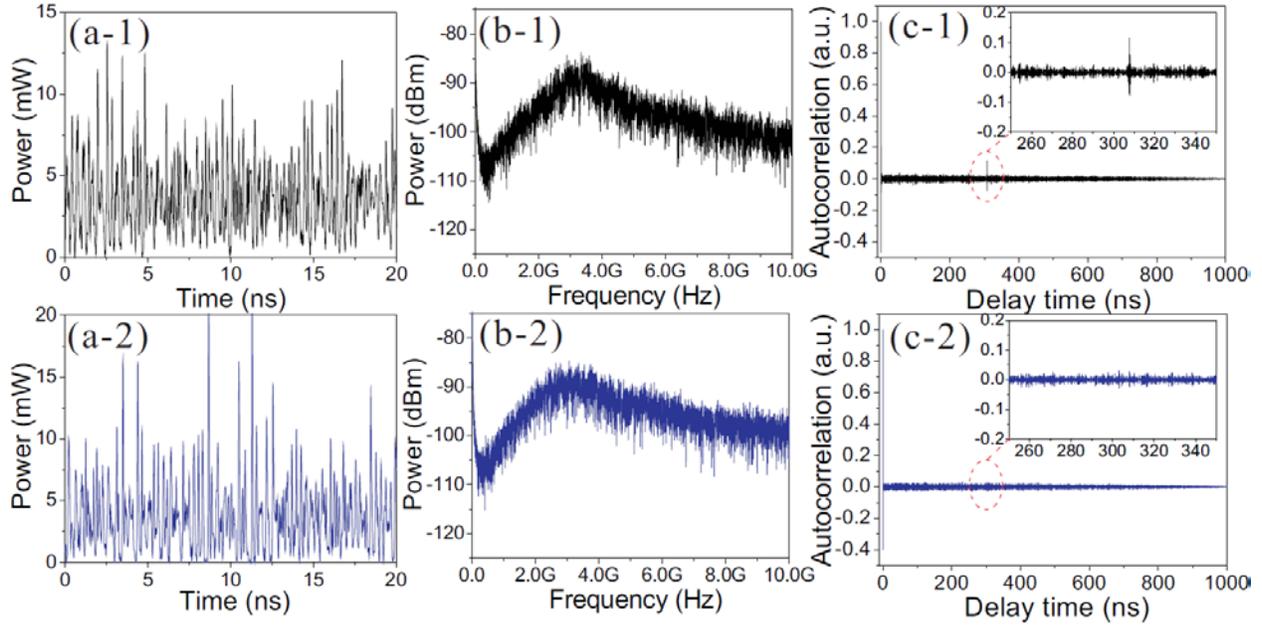

Figure 3. Simulation results of (a) time domain series, (b) power spectra, and (c) ACFs for laser diode subjected to single reflection feedback (1) and random distributed feedback (2), respectively.

Fig. 3(a)-(c) plot the time domain series of the chaotic laser output, power spectra, and ACFs of the output signals for single reflection feedback and random distributed feedback, respectively. The total feedback reflectivity from the distributed feedback in the simulation is the same as the single reflection feedback, which is 0.09 in amplitude. With this level of feedback strength, the laser diode is perturbed into a state of chaotic emission as shown in column 1 of Fig.3. Although the power spectra for both cases look similar, the ACFs of the time domain series show obvious TDS for the single reflection feedback while the TDS is significantly suppressed and hardly noticeable for the random distributed feedback. This indicates that random distributed feedback is able to obscure the periodicity in the time domain series by introducing many sets of the external cavity modes to the laser diode, leading to chaotic emission free from the TDS.

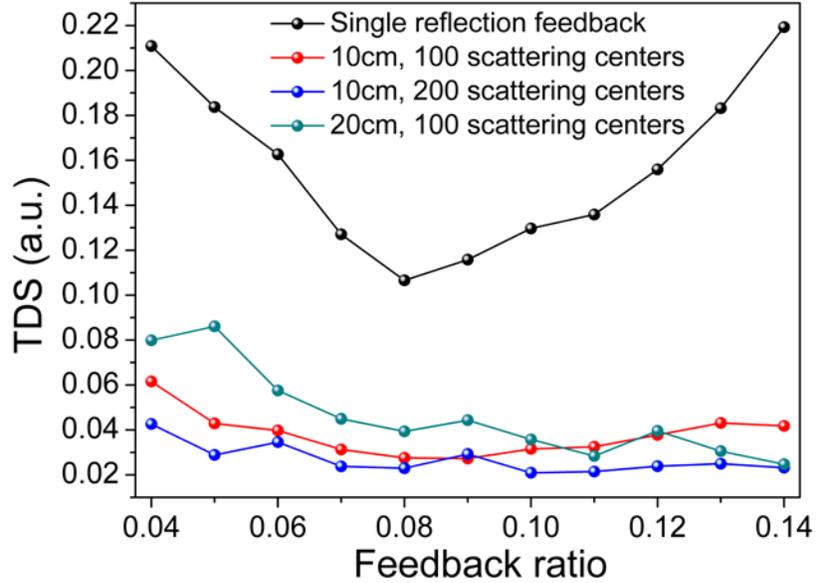

Figure 4. Computed TDS values of the chaotic laser subjected to single reflection feedback (black) and random distributed feedbacks from 10cm long random grating with 100 scattering centers (red), 10cm long random grating with 200 scattering centers (blue), 20cm long random grating with 100 scattering centers (dark cyan) as a function of feedback ratio.

The influence of reflection ratios on the TDS for both cases were also investigated as shown in Fig. 4. In the simulations, different distributed feedbacks were modeled in order to investigate the dependence of TDS suppression on the random grating parameters. The results show that whenever the grating length or the number of the scattering centers is changed, the significant suppression of the TDS could still be achieved with proper feedback ratio and a larger number of scattering centers is able to suppress the TDS more significantly. It is also found that over a relatively large range of reflection strengths, the random distributed feedback has much lower TDS values, with suppression ratios greater than 10 times that of the single reflection feedback. Furthermore, the TDS value of the random distributed feedback is less dependent on the feedback strength than that of the single reflection case.

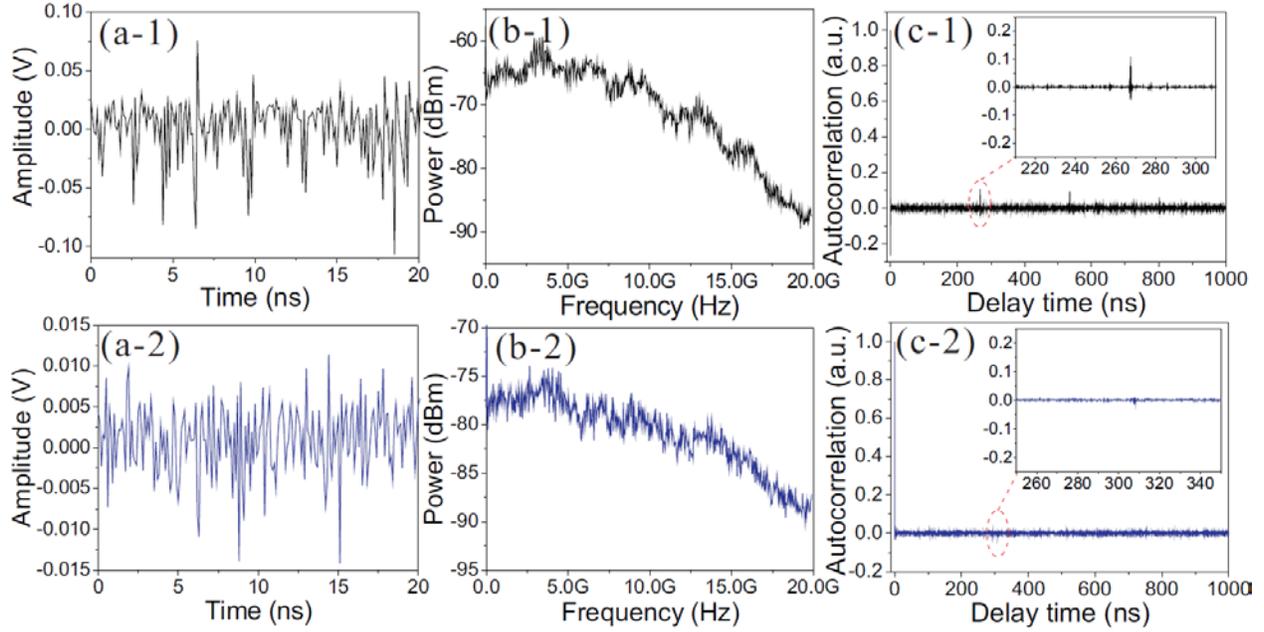

Figure 5. Experimental results of (a) time domain series, (b) power spectra, and (c) ACFs for laser diode subjected to single mirror feedback (1) and random grating feedback (2), respectively.

In experiments, a fiber random grating sample or a single mirror was connected with port 2 of OC 2 in Fig.1(a), respectively. The laser diode was operating with a driven current 2 times the threshold value, which emitted light with an output power around 1.0mW and an intrinsic relaxation oscillation frequency of 7GHz. The feedback from the random grating and the mirror was adjusted to 0.22 (in amplitude) and the experimental results are shown in Fig. 5. The measured time domain series, power spectra and ACFs of the chaotic output signal are obtained for both random grating feedback and mirror feedback. It is noted that the power spectra for both cases extend to lower frequency range, which could be explained by the anti-guidance effects in semiconductor lasers induced by the external feedback. The red-shift is even more obvious in the experimental results than within the simulation results due to the stronger feedback injected into the laser diode during the experiments. As expected, the ACF result for random grating feedback shows a significant suppression in the TDS and the minimum TDS value is obtained when the feedback strength is 0.22, which is only 0.0088 and almost negligible when compared with that of the mirror feedback result.

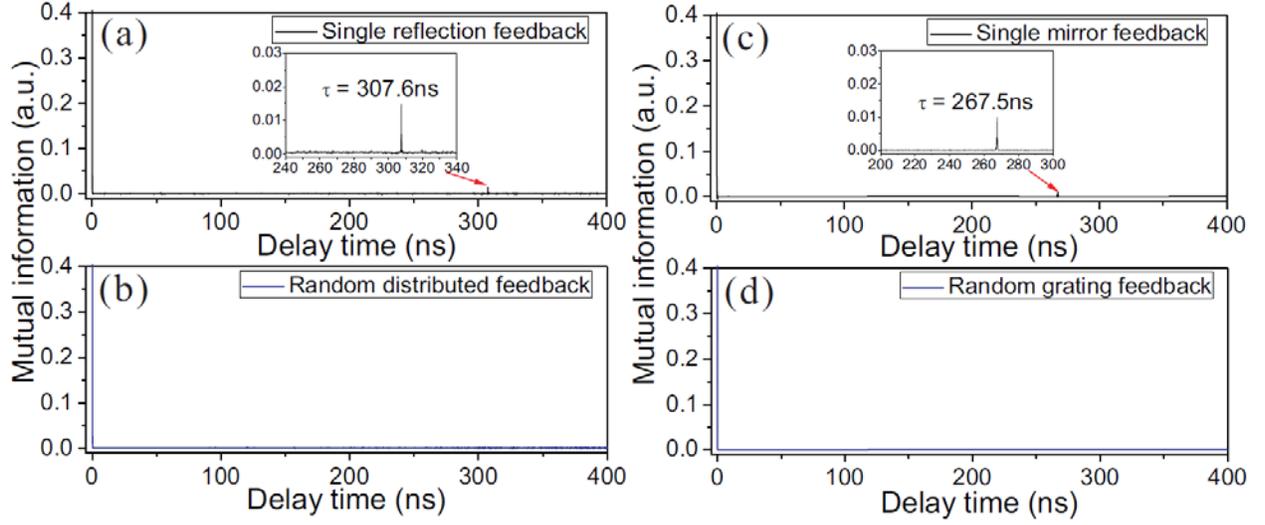

Figure 6. Mutual information of simulated results for chaotic laser subjected to (a) single reflection feedback and (b) random distributed feedback with a feedback ratio of 0.09; Mutual information of experimental results for chaotic laser subjected to (c) single mirror feedback and (d) fiber random grating feedback with a feedback ratio of 0.22.

Mutual information (MI) was also computed for the simulated and experimental chaotic time domain series to justify the suppression of TDS as shown in Fig. 6 using the following equation:

$$M(\tau) = \sum_{I(t),I(t+\tau)} p(I(t),I(t+\tau)) \log \frac{p(I(t),I(t+\tau))}{p(I(t))p(I(t+\tau))} \quad (4)$$

where $p(I(t), I(t+\tau))$ is the joint probability density, $p(I(t))$ and $p(I(t+\tau))$ are the marginal probability densities. It is clearly shown that the random distributed feedback from the fiber random grating is able to effectively cancel the TDS in the chaotic time domain series for both simulated and experimental results, which is mainly attributed to two effects: i) as the TDS in the single reflection scheme originates from the beating between the longitudinal modes in the single external cavity, the large number of scattering centers in the fiber random grating introduces large numbers of external cavity modes competing for the limited optical gain. Chaotic instability with high dimension is then expected and a strong and stable beating between any two of these longitudinal modes is prevented; ii) the random distributed feedback from the random grating as indicated in the noise-like interfering reflection spectrum due to the individual point writing technique with less phase control induces the frequency-dependent group delay, which corresponds to the group velocity dispersion, on different frequency components of the optical feedback to the laser diode, leading to the obscuring of the information of round trip time delay.

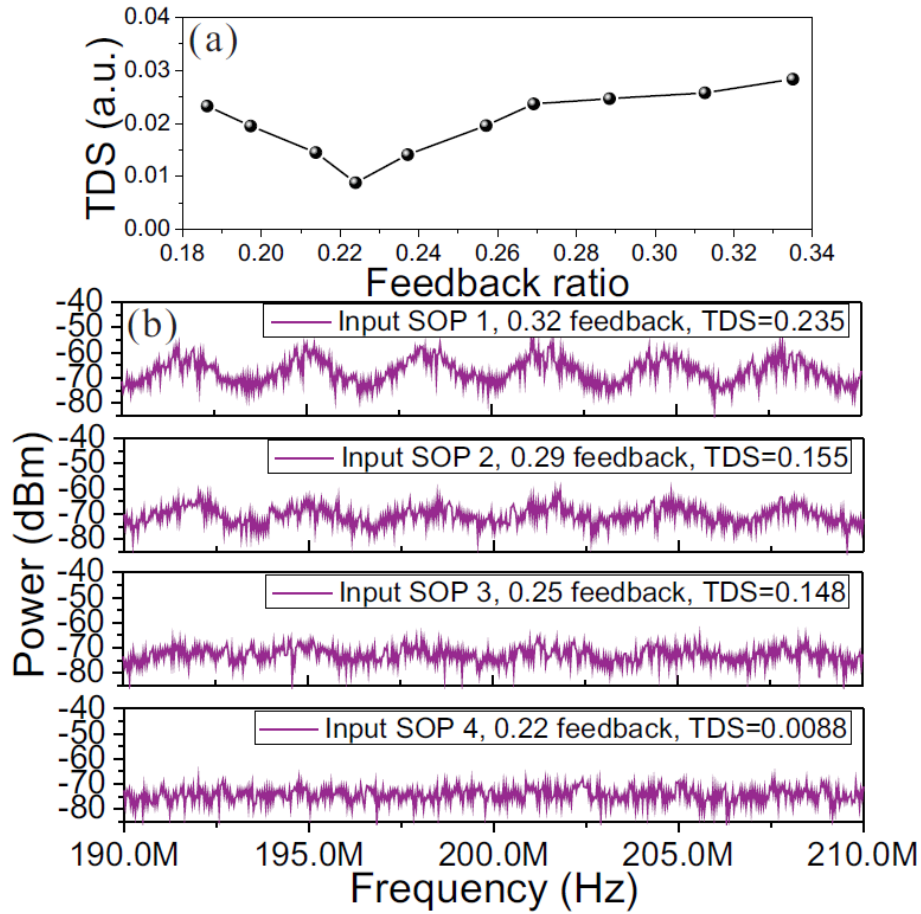

Figure 7. (a) TDS value of the chaotic laser subjected to the random grating feedback with the optimized input SOP to the fiber random grating sample; (b) Power spectra of the chaotic laser output with different input SOPs to the fiber random grating sample and the corresponding TDS values.

Due to the randomly distributed index modification spots along the fiber random grating, the directions for the principal birefringence axes of each scattering center are different from each other. Thus it is impossible to realize the alignment of the SOP of the light launched into the random grating with the directions of the local principal birefringence axes at each scattering center, resulting in different reflection coefficients at each reflector. The suppression performance of the random distributed feedback therefore depends on the SOP of the light that was launched into the random grating sample, which could be studied by adjusting PC 1 and keeping PC 2 fixed. It is found that the SOP of the light launched into the random grating significantly influences the TDS of the chaotic output. By adjusting the SOP of the input light to an optimized state where a large number of scattering centers have almost comparable contributions to the reflection, the dependence on the feedback strength of the TDS for the random grating feedback is obtained as shown in Fig. 7(a). The feedback strength was varied through the optical attenuator in Fig. 1(a) and both PCs were kept unchanged. The measured curve has a similar trend with the simulated result shown in Fig. 4 and shows low TDS values over the measured feedback strength range as expected. However, when the SOP of the light launched into the random grating was changed, the feedback strength as well as the TDS experiences large variations as shown in Fig. 7(b). This phenomenon is mainly attributed to the polarization dependent scattering from the scattering centers that are fabricated through the fs laser modification in the random grating. The resultant random birefringence makes the scattering centers highly polarization dependent and scatter the input light at different reflection levels according to the polarization matching conditions. Consequently, two distinct situations are expected: i) it is possible that there is one or two scattering centers that scatter back dominant light power over the others when perfectly polarization-matched with that of the input light, which renders similar results with the single reflection feedback and large TDS values; ii) On the other hand, it is also possible to find an SOP of the input light which leads to a light backscattering almost equally shared among a number of scattering centers rather than a reflection by one or two dominant scattering centers, which effectively takes advantages of the random distributed feedback from the random grating and obscures the periodicity of the time series. The first three power spectra in Fig. 7(b) belonging to the first situation clearly show a periodicity with a frequency spacing of 3.25MHz, which corresponds to the total ring length of 61.52m. While the periodicity in the last power spectrum which belongs to the second situation is hardly observed. Naturally, the TDS values in the first situation are much larger than that in the second situation.

In conclusion, the suppression of TDS has been theoretically and experimentally realized in a semiconductor laser subjected to the distributed feedback from a fiber random grating. A theoretical model is established for numerical simulations, which predicts the robustness of the random distributed feedback in eliminating the periodicity in laser diode-based optical chaos. The experimental results show a tremendous suppression in TDS with a minimum value of 0.0088 to date, allowing the effective and simple approach applicable in the concealment of the encryption system parameters and hence improving the security in optical chaos based communications.

The authors are thankful to the Natural Sciences and Engineering Research Council of Canada (NSERC) Discovery Grant and Canada Research Chair Program (CRC in Fiber Optics and Photonics).## References

1. Feigenbaum, M. J. The onset spectrum of turbulence. *Physics Letters A*, 74, 375-378 (1979).
2. Henry, C. Theory of the linewidth of semiconductor lasers. *IEEE Journal of Quantum Electronics*, 18, 259-264 (1982).
3. Ohtsubo, J. Semiconductor lasers: stability, instability and chaos (Vol. 111). Springer (2012).13